# DEVELOPMENT OF AN INTERPRETER PROGRAM FOR AN EQUIPMENT CONTROL

Y. Furukawa, M. Ishii, T. Nakatani, T. Ohata
SPring-8, Hyogo 679-5198, Japan

*Abstract*

A message oriented software scheme is used for the SPring-8 beamline control [1]. In the early stage, we developed customized control programs for each beamline. As the number of beamlines increased, however, it became necessary to introduce a general software scheme that could handle the control sequence for every beamline. We developed a new server program, Command Interpreter (CI), as a software framework. The CI interprets a high-level compound message issued from client programs and decomposes it to a set of primitive control messages. The messages are sent to the VME computers afterwards. For flexibility, the operation sequences specific to the individual beamline component are standardized and defined in the interface definition files. Using the CI, the response time overhead is reduced to 26% and the XAFS measurement time also decreases by about 50%.

## 1 INTRODUCTION

SPring-8 was successfully dedicated to public use on October 1997. The beamline control software of the SPring-8 is based on the same framework of the SPring-8 storage ring control software architecture [1]. A control message is sent from X11 based graphical user interface (GUI) programs to a message server (MS). The MS forwards the message to equipment managers (EM). These processes communicate by exchanging the messages formatted as "S/V/O/C" [2]. The EM manages only individual I/O channels such as stepper motors, analog I/Os and digital I/Os. Complex sequence operations, such as the correlated stepper motor operation of the standard X-ray monochromator [3], are achieved by GUI built-in sequence routines. Users conducting experiments should access the beamline control system not only via the local GUI terminal but also via remote computers for their experiments. The GUI process receives a message from the user's computer in order to change the X-ray energy by moving the monochromator for example. The GUI accepts a single message or compound message from the user's computers.

Optical components, such as a monochromator and mirrors, have been standardized for the public beamlines. The control sequences and parameters were built-in software code in the GUI programs. However, a more flexible framework was required to maintain not only the control parameters but also the sequences for the optical component especially for the contracted beamlines, because non standard components have been introduced.

The GUI programs communicate with the MS with one message, i.e. the GUI program sends a message then waits for a return message. This ensures the result of the request message, but the GUI can not respond to an X event or user messages while waiting for a response message from the EM.

As a result of this design, the response of the GUI to the user's computer was fairly slow in the case of spectroscopy beamlines, such as X-ray absorption fine spectrum (XAFS) beamlines.

The command interpreter is introduced to the SPring-8 beamline control system to improve the response time of the control system and to introduce flexibility.

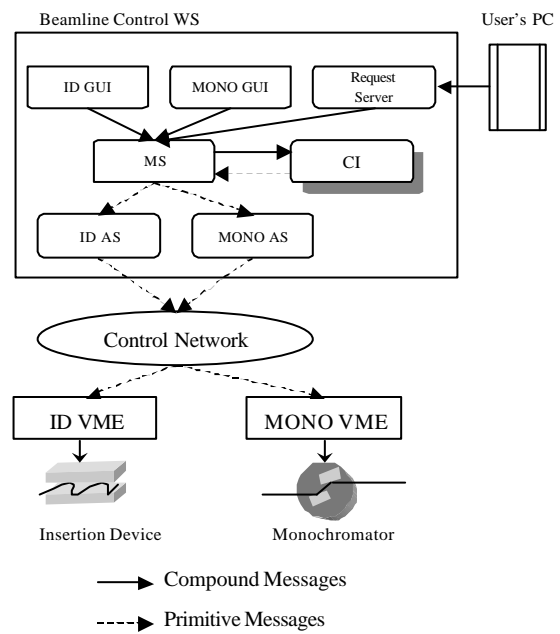

Figure 1: Message flow diagram.

## 2 DESIGN OF THE COMMAND INTERPRETER

### 2.1 Conceptual Design

The Command Interpreter (CI) is designed as a server process that communicates with the MS only. The internal structure of the CI is message driven. The CI accepts a compound message from client processes such as a GUI process or remote user's computer and interprets the message into primitive messages which the EM can accept. The primitive messages are sent to the MS at first, then to the EM (Figure 1). The CI does not wait for the returned message from the EM after it sent primitive messages. The EM executes the primitive message to move a stepper motor for example.

Information peculiar to the individual component is written in an interface definition file ("configuration file") described in the following subsection.

For example, one can control a monochromator by sending a message to the CI to change a Bragg angle, X-ray energy or X-ray wavelength. Referencing the configuration file for the monochromator, the CI interprets the message and calculates the pulse position of the stepper motors. Then the CI sends primitive messages to the EM.

### 2.2 Configuration Files

The configuration file contains component specific information. A simplified configuration file for the standard monochromator is shown in Figure 2 for example.

The configuration file consists of one or more "sections". Each section contains a definition of a high-level object such as monochromator or mirror. A section consists of the object name definition entry (line 2 in Figure 2) and subsections. Each subsection has entries, each entry consists of the entry name and following parameter(s). In a "subsection object", low-level objects that the EM can accept are defined (lines 48). In a "subsection function" (line 27-33), how the CI interprets the compound message and decomposes it into the primitive messages are defined.

### 2.3 Example of Message Flow

As an example, one can send a compound message "S/put/bl_01b1_mono/5.6degree" to the CI in order to set the Bragg angle of the monochromator. The CI interprets the message and decomposes it into primitive messages, "S/put/bl_01b1_mono_theta/10800pulse" and "S/put/bl_01b1_mono_y1/-153715pulse".

To obtain the present Bragg angle, one can send a compound message "S/get/bl_01b1_mono/angle" to the CI. The CI sends a primitive message

```
1   SECTION OBJECT
2     objectname bl_01b1_mono
3
4     SUBSECTION OBJECT
5       objectname theta
6       sendcomplement position
7     ENDSUBSECTION
     ..........
15    SUBSECTION CONT
16      constname y1_coeff
17      type double
18      value 1000.0
19    ENDSUBSECTION
     ..........
27    SUBSECTION FUNCTION
28      apply put/%fdegree
29      range 3.0 27.0
30
31      function theta [target()*theta_coeff] pulse
32      function y1 [-15.0/sin(target())*y1_coeff] pulse
33    ENDSUBSECTION
34
35    SUBSECTION FUNCTION
36      apply get/angle
37
38      function [theta/theta_coeff] degree
39    ENDSUBSECTION
40  ENDSECTION
```

Figure 2: Example of configuration for the monochromator.

"S/get/bl_01b1_mono_theta/position" then obtains the reply "S/bl_01b1_mono_theta/get/S/10800pulse". Evaluating the line 38, the CI returns a reply "S/bl_01b1_mono/get/S/5.6degree".

Figure 3 shows a message flow diagram as a comparison of conditions before introducing and after. The CI is designed to send primitive messages to move the low-level objects without waiting for returned messages from the EM. The CI is also designed to obtain the position of the low-level objects periodically when they are moving. Thus the CI can respond immediately to the client message whenever the client asks the the monochromator for the current angle. As seen from Figure 1, the response time to the client is reduced, and the start time difference between the theta1 and y1 axes is shortened.

## 3 APPLICATIONS

The SPring-8 standard monochromator, based on a double crystal monochromator, is controlled under two axis correlation mode. One is for the Bragg angle control and the other for the first crystal position control [3]. These two axes are driven by stepper motors. The configuration file for the standard monochromator is longer than that shown in Figure 2, because it needs additional subsections for the messages to obtain the current X-ray energy and wavelength. However the length of the configuration file is much shorter and simpler than if it were written in C-language, i.e. the number of lines is reduced to about 1/5.

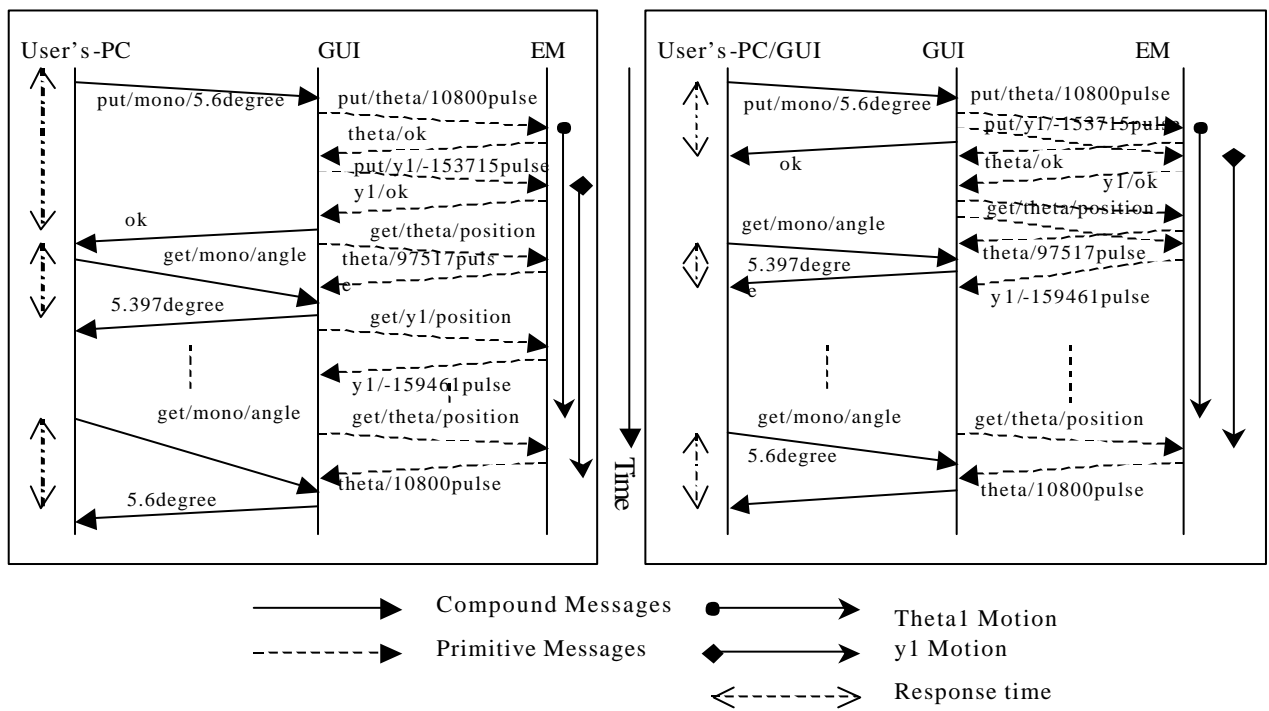

Figure 3: Message flow diagrams. Left: before introducing the CI, right: after introducing the CI.

There are many kinds of mirrors in the SPring-8. The CI laps these differences and the clients' software can control every mirrors with a same manner. We can adapt the CI to the new type mirror by slight modification of the configuration file.

## 4 PERFORMANCE IMPROVEMENT

The CI was initially applied in the monochromator of BL15XU, a contract beamline of the Advanced Materials Laboratory. The BL15XU monochromator is a non-standard monochromator in the SPring-8 because it covers a wider X-ray energy region. The CI laps the differences of the monochromator from clients' software. We made minor adjustments to the GUI programs to enable the standard monochromator to be used for the BL15XU. Before introducing the CI, it took about 30 minutes to measure a Cu-K XAFS spectrum (8970eV to 9010eV by 0.2eV step). After introducing the CI, the time was reduced to 15 minutes.

Control of BL01B1, a XAFS beamline, has also been upgraded using the CI. The CI controls a monochromator, X-ray mirrors and slits. The time needed to move the monochromator Bragg angle from 8.0 degrees to 8.1 degrees in 0.001 steps was measured on an HP/B2000 workstation with an HP-UX 10.20. The CI improved the moving time from 80sec to 46sec. The total measurement time, the sum of X-ray intensity measurement time and the moving angle time, was improved from 180sec to 146sec.

We also measured response time to the remote user's computer. The mean response time to get the Bragg angle of the monochromator was reduced from about 150msec to 40msec.

## 5 CONCLUSION

The CI was successfully introduced into the SPring-8 beamline control system. It made installation of the new component to the system easier and improved measurement time. Seven beamlines are already being operated using the CI. We plan to upgrade the control software using the CI for the existing beamlines and apply it to the newly constructed beamlines.

The CI is flexible and can be applied to other equipment controls, such as diffractometers, goniometers and detectors.

## REFERENCES

[1] T. Ohata *et al.*, "SPring-8 Beamline Control System", ICALEPCS'99, Trieste Italy, (1999) p.666.
[2] R. Tanaka *et al.*, "Control System of the SPring-8 Storage Ring", ICALEPCS'95, Chicago, USA, (1995) p.201 .
[3] M. Yabashi *et al.*, "SPring-8 standard x-ray monochromator", Proc. SPIE 3773, 19 July 1999, Denver, Colorado, U.S. (1999) 2.